\documentclass{PoS}
\usepackage{amsmath}

\usepackage[utf8]{inputenc}

\title{1-Loop Matching of gauge invariant dim-6 operators for \textit{B} decays}

\ShortTitle{1-Loop Matching of gauge invariant dim-6 operators for $B$ decays}

\author{\speaker{Jason Aebischer}\\
        Albert Einstein Center for Fundamental Physics, Institute
  for Theoretical Physics,\\ University of Bern, CH-3012 Bern,
  Switzerland.\\
        E-mail: \email{aebischer@itp.unibe.ch}}

\author{Andreas Crivellin\\Paul Scherrer Institut, CH--5232 Villigen PSI, Switzerland\\
        E-mail: \email{andreas.crivellin@cern.ch}}

\author{Matteo Fael\\
        Albert Einstein Center for Fundamental Physics, Institute
  for Theoretical Physics,\\ University of Bern, CH-3012 Bern,
  Switzerland.\\
        E-mail: \email{fael@itp.unibe.ch}}

\author{Christoph Greub\\
        Albert Einstein Center for Fundamental Physics, Institute
  for Theoretical Physics,\\ University of Bern, CH-3012 Bern,
  Switzerland.\\
        E-mail: \email{greub@itp.unibe.ch}}

\abstract{Physics beyond the Standard Model, realized above the electroweak scale, can be incorporated in a model independent way in the Wilson coefficients of higher dimensional gauge invariant operators. 
In these proceedings we review the matching of the $SU(3)_C\times SU(2)_L\times U(1)_Y$ gauge invariant dimension-six operators on the effective Hamiltonian governing $b\to s$ and $b\to c$ transitions, including the leading 1-loop effects~\cite{Aebischer:2015fzz}.}

\FullConference{16th International Conference on B-Physics at Frontier Machines\\
                 2-6 May 2016\\
                 Marseille, France}

\begin{document}

\section{Introduction}

Despite numerous confirmations of its validity, the Standard Model (SM) of particle physics is thought to be only an effective theory valid up to a new physics scale $\Lambda$, where additional dynamical degrees of freedom enter. The SM effective theory (SMET) Lagrangian can be written in the following form \cite{Buchmuller:1985jz,Grzadkowski:2010es}:
\begin{equation}\label{eq:lag}
 \mathcal{L}_{SM} =   \mathcal{L}_{SM}^{(4)}   + \frac{1}{\Lambda }  C_{\nu\nu}^{(5)} Q_{\nu\nu}^{(5)} 
  + \frac{1}{\Lambda^2} \sum_{k} C_k^{(6)} Q_k^{(6)}  + \mathcal{O}\left(\frac{1}{\Lambda^3}\right)\,.
\end{equation}
In this approach, physics beyond the SM is encoded in a model independent way in the Wilson coefficients of the higher dimensional operators $Q_k$. For $B$~physics, the Wilson coefficients $C_k^{(6)}$, multiplying the dimension-six operators are relevant, while the dimension-five Weinberg operator only provides neutrino mass terms after electroweak (EW) symmetry breaking~\cite{Weinberg:1979sa}. 

\begin{figure}[h]
  \centering
  \includegraphics[width=\textwidth]{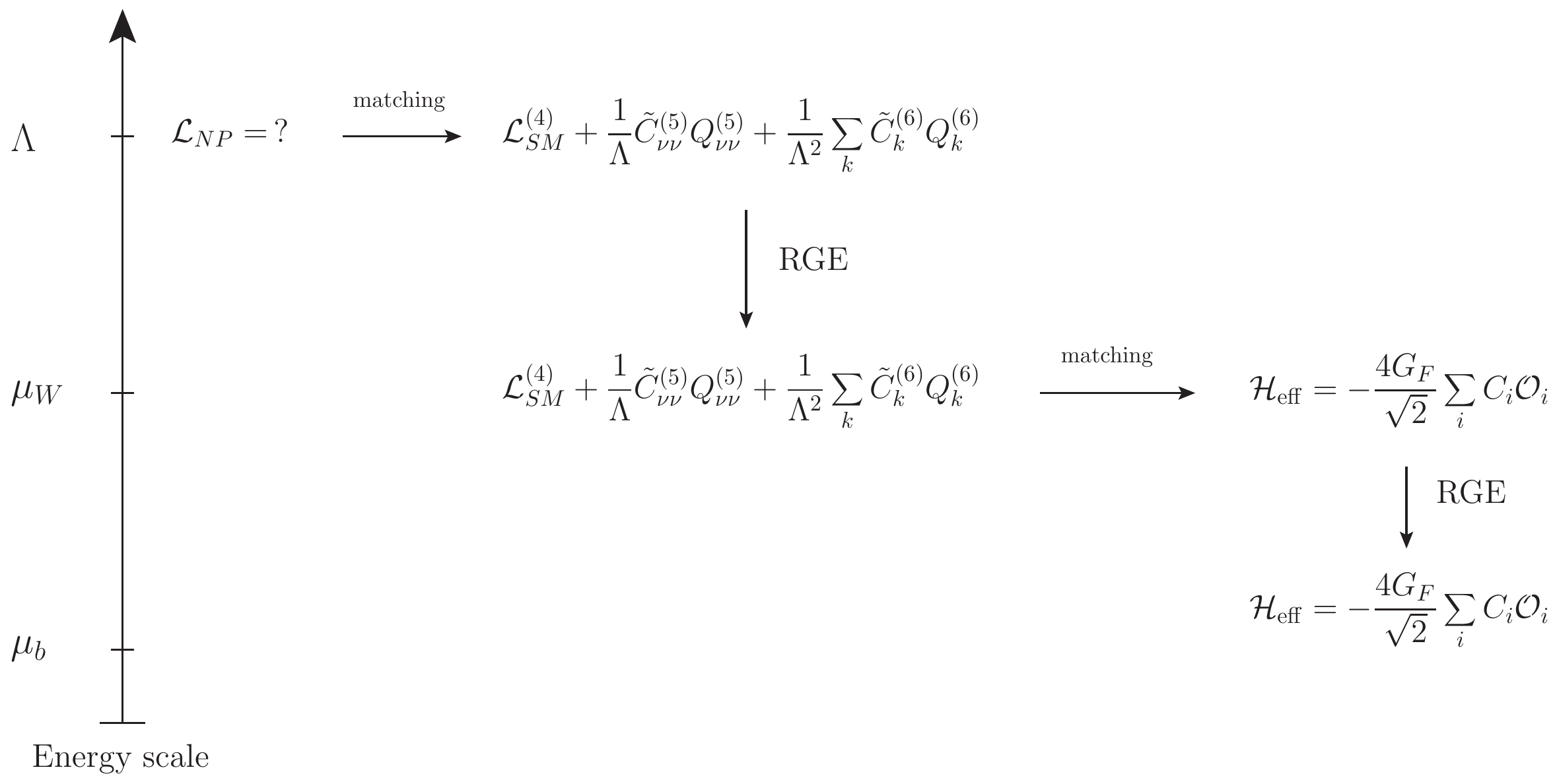}
  \caption{Mass scale hierarchy: 1) Matching of NP model onto SMET at high scale $\Lambda$. 2) RGE evolution down to EW scale $\mu_W$. 3) Matching of dimension-six operators on effective $B$~physics Hamiltonian. 4) RGE evolution down to $B$~scale $\mu_b$.}
  \label{fig:massscale}
\end{figure}

In order to compare the predication of a NP model to $B$ physics observables, the following steps have to be performed. 
\begin{enumerate}
	\item Running of the Wilson coefficients $C_k^{(6)}$ from the matching scale $\Lambda$ to the electroweak (EW) symmetry breaking scale $\mu_W$~\cite{Jenkins:2013zja}.
	\vspace{-2mm}
	\item EW symmetry breaking is performed and the SMET Lagrangian is matched onto the effective Hamiltonian governing $B$~physics.\footnote{For the corresponding calculation in the lepton sector see Refs.~\cite{Crivellin:2013hpa,Pruna:2014asa}.}
	\vspace{-2mm}
	\item Renormalization group equations can be used to perform the evolution of the Wilson coefficients from the electroweak scale down to the $B$~scale $\mu_b$ (see for example~\cite{Borzumati:1999qt,Buras:2000if}).
\end{enumerate}
This procedure is depicted in Fig.~\ref{fig:massscale}. The requirement of gauge invariance reduces the number of operators in $B$ physics~\cite{Alonso:2014csa} and correlates charged with neutral currents (see for example~\cite{Calibbi:2015kma}).

Above the EW symmetry breaking scale the gauge invariant dimension-six operators are given in the interaction basis, since the mass basis it not defined above $\mu_W$. After the EW symmetry breaking, the fermions are rotated into the mass eigenstates by diagonalizing their mass matrices, which affects the Wilson coefficients. All rotation matrices appearing in the operators can be absorbed by a redefinition of the Wilson coefficients, with the exception of the misalignment between the left-handed up-quark and down-quark rotations, i.e.\ the Cabibbo-Kobayashi-Maskawa matrix.
\medskip

In Ref.~\cite{Aebischer:2015fzz} we performed the matching of the SMEFT onto the effective Hamiltonian of $B$~physics by integrating out all heavy degrees of freedom (compared to $\mu_b$), i.e. the top quark, the $W$ and $Z$ bosons and the Higgs field. The full tree-level matching has been computed for $b\rightarrow s$ and $b\rightarrow c$ transitions. In addition, 1-loop contributions have been performed, which involve dimension-six operators that do not enter the matching at tree level.

\section{1-loop matching}

Operators with a top quark do not contribute to the $b \to s$ and $b \to c$ transitions at the tree-level, since the top is not contained in the $B$ physics Hamiltonian. The 1-loop matching contributions due to dimension-six operators 
containing right-handed top quarks can be divided in the following six 
classes: 
\begin{enumerate}
  \item 4-fermion operators to 4-fermion operators ($\Delta B =\Delta S =1$).
	\vspace{-2mm}
  \item 4-fermion operators to 4-fermion operators ($\Delta B =\Delta S =2$).
	\vspace{-2mm}
  \item 4-fermion operators to $\mathcal{O}_7$ and $\mathcal{O}_8$.
	\vspace{-2mm}
  \item Right-handed $Z$ couplings to $\mathcal{O}_9$, $\mathcal{O}_{10}$ and $\mathcal{O}^q_{3-6}$.
	\vspace{-2mm}
  \item Right-handed $W$ couplings to $\mathcal{O}_7$ and $\mathcal{O}_8$.
	\vspace{-2mm}
  \item Magnetic operators to $\mathcal{O}_7$, $\mathcal{O}_8$, $\mathcal{O}_9$, $\mathcal{O}_{10}$ and $\mathcal{O}_4^q$.
\end{enumerate}

As an example we consider the dimension-six operator $Q_{\varphi ud}=(\tilde{\varphi}^\dagger \, i  D_\mu\varphi )( \bar{u}_i \gamma^\mu P_R d_j )$, which couples the $W$-boson to right-handed quarks. This anomalous $W-t-b$ coupling induces a non-zero contribution to the magnetic operators $\mathcal{O}_7, \mathcal{O}_8$. The magnetic operators which are contained in the $\Delta B=\Delta S=1$ effective Hamiltonian read:
\begin{align}
  \mathcal{O}_7=\frac{e}{16\pi^2}m_b(\bar{s}\sigma^{\mu\nu}P_Rb)F_{\mu\nu}\,,\qquad
 \mathcal{O}_8=\frac{g_s}{16\pi^2}m_b(\bar{s}\sigma^{\mu\nu}P_RT^Ab)G_{\mu\nu}^A\,.
\end{align}
The matching contributions from $Q_{\varphi ud}$ are given by (in agreement with ~\cite{Grzadkowski:2008mf,Drobnak:2011aa}):

\begin{align}
   C_7 =   \frac{m_t}{m_b} \,   
   \frac{v^2}{\Lambda^2}   E_{ \varphi ud}^7 (x_t)  \,   
   \tilde{C}_{\varphi ud}^{33} \,   V_{ts}^*\,,\qquad
   C_8 &=   \frac{m_t}{m_b}  
   \frac{v^2}{\Lambda^2}     E_{ \varphi ud}^8 (x_t)  \,   
   \tilde{C}_{\varphi ud}^{ 33} \,   V_{ts}^*\,,\label{eqn:C8phiud}
\end{align}
where the dimensionless $x_t=m_t^2/M_W^2$-functions are defined in Ref.~\cite{Aebischer:2015fzz}.

\section{Conclusions}
We presented the complete tree-level matching coefficients for $b\to s$ and $b\to c$ transitions including lepton flavor violating operators. 27 out of the 59 gauge invariant dimension-six operators contribute to the tree-level matching. Another 14 operators enter the 1-loop matching. They involve 4-fermion operators, electromagnetic and chromomagnetic dipole operators as well as operators involving Higgs and quark fields. 
Once the running from the EW scale down to the $B$ meson scale will be performed, our results can be used to perform systematic tests on the sensitivity of $B$ physics observables on the dimension-six operators. 

\small
\bibliographystyle{unsrt}
\bibliography{sample} 

\end{document}